\documentclass[]{jfm}

\usepackage{natbib}
\usepackage{hyperref}
\usepackage{pgfplots}

\usepackage{graphicx}
\usepackage{amsmath}
\usepackage{newtxtext,newtxmath}

\newcommand{\aver}[1]{ \! \left\langle {#1} \right \rangle \!}

\newcommand{\DR}{\mathcal{DR}}

\title[Turbulent drag reduction in compressible flows]{Turbulent drag reduction with streamwise travelling waves in the compressible regime}

\author{Federica Gattere\aff{1}, Massimo Zanolini\aff{1}, Davide Gatti\aff{2}, Matteo Bernardini\aff{3} and Maurizio Quadrio\aff{1}}
\affiliation{\aff{1} Dipartimento di Scienze e Tecnologie Aerospaziali, Politecnico di Milano,
via La Masa 34, 20156 Milano, Italy \\
\aff{2} Institute for Fluid Mechanics, Karlsruhe Institute of Technology, Kaiserstr. 10, 76131 Karlsruhe, Germany \\
\aff{3} Department of Mechanical and Aerospace Engineering, Sapienza University of Rome, via Eudossiana 18, 00184 Rome, Italy
}

\begin{document}
\maketitle

\begin{abstract}
The ability of streamwise-travelling waves of spanwise velocity to reduce the turbulent skin friction drag is assessed in the compressible regime. Direct numerical simulations are carried out to compare drag reduction in subsonic, transonic and supersonic channel flows.
Compressibility improves the benefits of the travelling waves, in a way that depends on the control parameters: drag reduction becomes larger than the incompressible one for small frequencies and wavenumbers.
However, the improvement depends on the specific procedure employed for comparison. 
When the Mach number is varied and, at the same time, wall friction is changed by the control, the bulk temperature in the flow can either evolve freely in time until the aerodynamic heating balances the heat flux at the walls, or be constrained such that a fixed percentage of kinetic energy is transformed into thermal energy. 
Physical arguments suggest that, in the present context, the latter approach should be preferred. Not only it provides a test condition in which the wall-normal temperature profile more realistically mimics that in an external flow, but also leads to a much better scaling of the results, over both the Mach number and the control parameters. Under this comparison, drag reduction is only marginally improved by compressibility.
\end{abstract}

\section{Introduction}
\label{sec:introduction}

One of the distinctive features of fluid turbulence is the ability to transport and mix mass and momentum more effectively than a laminar flow, resulting in more intense wall shear stress and a larger friction drag \citep{fukagata-iwamoto-kasagi-2002}. 
Flow control for skin-friction drag reduction aims to mitigate the negative effects of turbulence near the wall, in order to cut energy consumption and to improve cost effectiveness and environmental footprint. 
This is of particular interest in aeronautics: nearly 50\% of the total drag of a civil aircraft is due to the viscous drag caused by the interaction of the turbulent boundary layer with the surface \citep{gadelhak-pollard-1998}. 
An efficient drag reduction technology capable to achieve even a tiny drag reduction rate would yield enormous economic and environmental benefits.

Drag reduction strategies are often classified as passive or active. The former do not require extra energy, and usually exploit a non-planar wall \citep[see][for an exception]{foggirota-etal-2023-2}. Among them, riblets \citep{bechert-etal-1997} are the closest to be implemented in practical applications. 
Laboratory tests show that they can reduce drag up to 8--10\% at low Reynolds numbers; on considering their requirement of periodical maintenance, though, riblets do not yield enough economical benefits to be routinely used yet.
Active strategies, instead, require actuation, and external energy to work. 
Those involving the motion of the wall are an interesting category, and include spanwise wall oscillations \citep{jung-mangiavacchi-akhavan-1992}, streamwise-travelling waves of spanwise velocity \citep{quadrio-ricco-viotti-2009}, spanwise-travelling waves of spanwise velocity \citep{du-symeonidis-karniadakis-2002} and streamwise-travelling waves of wall deformation \citep{nakanishi-mamori-fukagata-2012}. 
They are all predetermined strategies, since the control parameters are set \textit{a priori}, and enjoy the relative simplicity resulting from the lack of sensors and feedback laws. However, several of them do not yield an energetic benefit once the control energy is accounted for.
This work focuses on the streamwise-travelling waves (StTW) of spanwise velocity introduced by \cite{quadrio-ricco-viotti-2009}. StTW are among the most promising techniques, because of their rather large net savings. This type of forcing, thoroughly reviewed by \cite{ricco-skote-leschziner-2021}, is defined by the following space-time distribution of the spanwise velocity component at the wall:
\begin{equation}
W(x,t) = A \sin(\kappa_x x - \omega t)
\label{eq:sttw}
\end{equation}
where $x$ and $t$ are the streamwise direction and time, $A$ is the forcing amplitude, $\kappa_x$ is the wavenumber and $\omega$ is the frequency (which define the wavelength $\lambda_x = 2 \pi/\kappa_x$ and the oscillation period $T=2\pi/\omega$). 
The spatially uniform spanwise-oscillating wall \citep{jung-mangiavacchi-akhavan-1992} and the stationary wave \citep{quadrio-viotti-luchini-2007, viotti-quadrio-luchini-2009} are two limit cases of the general forcing \eqref{eq:sttw}, obtained for $\kappa_x=0$ and $\omega=0$ respectively.

Via a generalized Stokes layer \citep{quadrio-ricco-2011}, StTW create an unsteady near-wall transverse shear which continuously changes the inclination of the near-wall structures in wall-parallel planes, weakening the regeneration mechanism of the near-wall cycle \citep{schoppa-hussain-2002}. 
Once actuation parameters are properly tuned, this process can even lead to the complete suppression of turbulence.

The spatially-uniform wall oscillation, studied in depth by \cite{quadrio-ricco-2004} in an incompressible channel flow at a Reynolds number (based on the friction velocity $u_\tau$, the fluid kinematic viscosity $\nu$ and the half-channel height) of $Re_\tau=200$, yields a drag reduction rate $\DR$ of 45\% (at $A^+ \equiv A/u_\tau = 12$) for the so-called `optimal' actuation period $T^+ \equiv T u^2_\tau / \nu \approx 100$. However, the maximum energy saving after the control energy is accounted for is found at lower forcing intensities, and amounts to 7\% only. 
The spatially-distributed StTW are a natural generalization of the wall oscillations, and present substantial advantages in terms of net savings. 
\cite{quadrio-ricco-viotti-2009} have shown how drag reduction, power input and total saved power vary with the control parameters. 
Depending on the ($\kappa_x, \omega$) value pair, drag increase or drag reduction can be achieved. The parameters yielding maximum drag reduction and maximum energy saving are almost coincident, and correspond (at this Reynolds number) to low frequencies and low wavenumbers. The largest drag reduction of 48\% (at $A^+=12$) still yields a positive net power saving of 17\%, and smaller forcing intensities lead to net savings as high as 32\%. StTW have been demonstrated in the lab with a pipe flow experiment \citep{auteri-etal-2010}, who measured up to 33\% drag reduction, and have been proven to work in boundary layers too \citep{skote-schlatter-wu-2015, bird-santer-morrison-2018}.

A number of practical aspects that need to be considered before declaring spanwise forcing as a viable strategy for applications has been recently considered. \cite{gatti-quadrio-2013, gatti-quadrio-2016} showed that the expected performance deterioration at larger Reynolds numbers, which afflicts all drag reduction strategies acting via near-wall turbulence manipulation, is only marginal for StTW and linked to the natural variation of the skin-friction coefficient itself with the Reynolds number. Once the performance of StTW is measured, as it should be, via the upward shift of the logarithmic portion of the mean velocity profile in the law-of-the-wall form, it becomes $Re$-independent, so that at flight Reynolds number 30\%--40\% friction drag reduction could be expected. \cite{marusic-etal-2021} hinted at an even better scenario for StTW at high $Re$, thanks to the interaction of the near-wall forcing with the large-scale outer motions of the turbulent boundary layer. \cite{banchetti-luchini-quadrio-2020} demonstrated the beneficial effect of skin-friction drag reduction via StTW on pressure drag when applied to bluff bodies of complex shape, and \cite{nguyen-ricco-pironti-2021} used spanwise forcing for separation control. 

One parameter that is crucial in aeronautical applications has received limited attention so far in drag reduction studies: the Mach number $M$, a parameter which quantifies the importance of compressibility effects. 
A few works, numerical \citep{duan-choudhari-2012, duan-choudhari-2014, mele-tognaccini-catalano-2016} and experimental, both in wind tunnel \citep{gaudet-1989, coustols-cousteix-1994} and with flight test \citep{zuniga-anderson-bertelrud-1992}, investigated the drag reduction effectiveness of riblets in a turbulent compressible boundary layer. 
Fewer studies have been carried out to assess how compressibility alters the drag reduction capabilities of active techniques: for example, \cite{chen-etal-2016} examined the uniform blowing or suction in an hypersonic turbulent boundary layer at free-stream Mach number of $6$.

As far as spanwise forcing goes, the large eddy simulation study of \cite{fang-lu-shao-2009} was the first to consider the spanwise oscillating wall in a turbulent channel flow at $M=0.5$, followed by the direct numerical simulation (DNS) study of \cite{ni-etal-2016} for a turbulent boundary layer at $M=2.5$. 
However, the first comprehensive study of compressibility effects in drag reduction via spanwise wall oscillations was performed by \cite{yao-hussain-2019}. They carried out DNS of a plane channel flow subjected to spanwise oscillating walls at $M = 0.3,0.8,1.5$, at $Re_\tau=200$, $A^+=12$ and $T^+$ in the range $25-300$. 
$\DR$ was found to be qualitatively similar to the incompressible case: for a given period $T^+$, $\DR$ increases with the amplitude $A^+$, at a rate that saturates when $A^+$ becomes large. 
For $A^+=12$, they reported $\DR$ increasing from $34.8 \%$ at $T^+=100$ for $M=0.3$ to an outstanding value of $47.1 \%$ at the largest period investigated $T^+=300$ for $M=1.5$. For $A^+=18$ and $M=1.5$, the flow reached relaminarization.
The effect of $Re$ was also investigated via a few additional cases run at $Re_\tau \approx 500$, confirming the related decline of $\DR$.
\cite{yao-hussain-2019} did not consider the impact of the Mach number on the power budget.
Both drag reduction and power budget performance were later discussed in the recent work by \cite{ruby-foysi-2022} for a channel flow at $M=0.3,1.5,3$ and $Re_\tau=200-1000$ forced by stationary waves with $A^+=12$ and $\kappa_x^+=0.0025-0.01$. They found the optimum $\kappa_x$ and the maximum net power saving to increase significantly with Mach, thus confirming the beneficial effect of compressibility.

When applying flow control for drag reduction in duct flows at various $M$, the thermodynamical properties of the flow change because of the increased bulk temperature, owing to the combination of the increased Mach number and the action of the control.
To understand whether changes of drag reduction with $M$ directly depend on compressibility, rather than indirectly deriving from temperature changes induced by changes of the skin friction drag, the comparison procedure between uncontrolled and controlled flows should decouple compressibility from purely thermodynamical effects. 
\cite{yao-hussain-2019} examined the effect of $M$ on $\DR$ by matching the semi-local Reynolds number (at half-channel height), which provides a relatively good collapse of $\DR$ between incompressible and compressible cases.
In the present work, we also propose a further, alternative approach: the value of the bulk temperature is constrained such that the amount of turbulent kinetic energy transformed into thermal energy remains constant, both across the variation of $M$ and between uncontrolled and controlled cases. 
This strategy presents a significant advantage. The simplified setup of the turbulent channel flow can be used in configurations where the coupling between the velocity and thermal fields is closer to that found in external flows, where the application of the spanwise forcing to reduce drag is more attractive. For example, compressible boundary layers of practical aeronautical interest are usually characterized by adiabatic or moderately cold walls, with a thermal stratification leading to a denser, colder outer region and a layer of warmer fluid in the near-wall zone.

The present work is the first comprehensive analysis of the StTW technique in the compressible regime. The only prior work is the single case computed by \cite{quadrio-etal-2022}, who studied by DNS the StTW applied on a portion of a wing in transonic flight at $M=0.7$ and $Re=3 \times 10^5$ (based on the free-stream velocity and the wing cord), finding that a localized actuation has the potential to boost the aerodynamic efficiency of the whole aircraft, with an estimate reduction of 9\% of the total drag of the airplane at a negligible energy cost.
In this work, we consider by DNS a compressible turbulent plane channel flow modified by StTW, and we aim at fully characterizing how $\DR$ and the power budget depend on the Mach number.

The paper is organized as follows. After this Introduction, Section \S\ref{sec:methods} describes the computational framework used to produce the DNS database, presenting the governing equations in \S\ref{subsec:equations}, the DNS solver in \S\ref{subsec:solver}, and the simulation parameters in \S\ref{subsec:parameters}.  \S\ref{subsec:performance} defines the parameters used to quantify drag reduction, and \S\ref{subsec:comparison} describes two approaches to compare unforced and forced compressible channel flows at different $M$.
In \S\ref{sec:results} the effects of the Mach number are discussed, first in terms of drag reduction in \S\ref{sec:DR}, and then in terms of power budgets in \S\ref{sec:power}. Lastly, in \S\ref{sec:conclusions} the main conclusions are briefly discussed. 

\section{Methods}
\label{sec:methods}

\subsection{Governing equations}
\label{subsec:equations}

The compressible Navier--Stokes equations for a perfect and heat-conducting gas are written in conservative form as:
\begin{equation}
\frac{\partial \rho}{\partial t} + \frac{\partial \rho u_i}{\partial x_i} = 0
\label{eq:mass}
\end{equation}
\begin{equation}
\frac{\partial \rho u_i}{\partial t} + \frac{\partial \rho u_i u_j }{\partial x_j} = - \frac{\partial p}{\partial x_i} +\frac{\partial \sigma_{ij}}{\partial x_j} + f \delta_{i1}
\label{eq:momentum}
\end{equation}
\begin{equation}
\frac{\partial \rho e }{\partial t} + \frac{\partial \rho(e+p/\rho) u_j }{\partial x_j} =  \frac{\partial \sigma_{ij} u_i}{\partial x_j} - 
\frac{\partial q_j}{\partial x_j} + f u_1 + \Phi.
\label{eq:energy}
\end{equation}
Here and throughout the paper, repeated indices imply summation; $\rho$ is the fluid density, $p$ is the pressure, $u_i$ is the velocity component in the $i$-$th$ directions, and $i = 1, 2, 3$ represent the streamwise ($x$), wall-normal ($y$) and spanwise ($z$) direction, respectively. 
The total energy per unit mass $e = c_{v} T + u_i u_i / 2$ is the sum of the internal energy and the kinetic energy, where $c_{v}$ is the specific heat at constant volume and $T$ the temperature. The viscous stress tensor $\sigma_{ij}$ for a Newtonian fluid subjected to the Stokes hypothesis becomes:
\begin{equation}
\sigma_{ij} = \mu \bigg( \frac{\partial u_i}{\partial x_j} + \frac{\partial u_j}{\partial x_i} - \frac{2}{3} \frac{\partial u_k}{\partial x_k} \delta_{ij} \bigg),
\label{eq:viscousstresstensor}
\end{equation}
where $\mu$ is the dynamic viscosity and $\delta_{ij}$ is Kronecker delta; the dependence of viscosity on the temperature is accounted for through the Sutherland's law. The heat flux vector $q_j$ is modelled after the Fourier law:
\begin{equation}
q_j = - k \frac{\partial T}{\partial x_j},
\label{eq:heatfluxvector}
\end{equation}
where $k = c_p \mu /Pr$ is the thermal conductivity, with $c_p$ the specific heat at constant pressure and $Pr$ the Prandtl number, set to $Pr= 0.72$. We consider the turbulent channel configuration, where the flow between two isothermal walls is driven in the streamwise direction by the time-dependent body force $f$ in Eq.\eqref{eq:momentum}, evaluated at each time step to maintain a constant mass flow-rate. The corresponding power is included in Eq.\eqref{eq:energy}, where the additional term
$\Phi$ represents a uniformly distributed heat source which controls the value of the bulk flow temperature \citep{yu-xu-pirozzoli-2019}.

\subsection{Solver}
\label{subsec:solver}

The flow solver employed for the analysis is STREAmS (Supersonic TuRbulEnt Accelerated Navier--Stokes Solver), a high-fidelity code designed for large-scale simulations of compressible turbulent wall-bounded flows that runs in parallel on both CPU or GPU architectures.

The code, developed by \cite{bernardini-etal-2021}, incorporates state-of-the-art numerical algorithms, specifically designed for the solution of compressible turbulent flows, with a focus on the high-speed regime.
The distinctive feature of the solver is the methodology adopted for the discretization of the convective terms of the Navier--Stokes equations with hybrid, high-order, energy-consistent/shock-capturing schemes in locally conservative form. 
An energy-preserving discretization, based on sixth-order central approximations, is applied where the solution is smooth, and guarantees discrete conservation of the total kinetic energy in the limit case of inviscid, low-speed flows. This is the case of interest for all the simulations presented in this study, where shock waves do not occur.
The Navier--Stokes equations are reduced to a semi-discrete system of ordinary differential equations,  integrated in time using a three-stages third-order Runge--Kutta scheme.
The solver is written in Fortran, and uses the MPI paradigm with a double domain decomposition; in its current version \citep{bernardini-etal-2023}, it can be run on modern HPC architectures based on GPU acceleration.
All the computations reported in this work have been performed using the CUDA Fortran backend, capable of taking advantage of the Volta NVIDIA GPUs available on Marconi 100 of the Italian CINECA supercomputing center.

\subsection{Parameters and computational setup}
\label{subsec:parameters}

A wall-bounded turbulent flow in the compressible regime is described by three independent parameters: the Reynolds number, the Mach number and a third parameter that specifies the thermal condition of the wall. For the channel flow configuration, relevant parameters are usually defined using bulk quantities, i.e. the bulk density $\rho_b$, the bulk velocity $U_b$ and the bulk temperature $T_b$:
\begin{equation}
 \rho_b = \frac{1}{2h} \int^h_{-h} {\aver{\rho} \, dy}, \quad
 U_b = \frac{1}{2h \rho_b} \int^h_{-h} {\aver{\rho u} \, dy}, \quad
 T_b = \frac{1}{2h\rho_b U_b} \int^h_{-h} {\aver{\rho u T} \, dy} .
\end{equation}
The operator $\aver{ \cdot }$ computes a mean value by averaging over time and homogeneous directions.

The main goal of this work is to understand the effect of Mach number. 
Since the control is wall-based and the control parameters are known \citep{gatti-quadrio-2016} to scale in viscous units, i.e. with the friction and density at the wall, it is convenient \citep{coleman-kim-moser-1995} to define the Mach number as $M_w^b = U_b / c_w$, in which the superscript and subscript emphasize that the velocity scale is $U_b$ and the speed of sound $c_w=\sqrt{\gamma R T_w}$ is evaluated at the (reference) wall temperature $T_w$.
Three sets of simulations are performed, at $M_w^b=0.3,0.8,1.5$. These values are identical to those used by \cite{yao-hussain-2019} in their study of the oscillating wall. 
The simulations are run at a constant flow rate or CFR \citep{quadrio-frohnapfel-hasegawa-2016}: the pressure gradient evolves in time to keep a constant $U_b$.
For all cases, the bulk Reynolds number $Re_b = \rho_b U_b h/ \mu_w$ is chosen in such a way that the corresponding friction Reynolds number is fixed to the target value for the uncontrolled simulations.
Although most of the incompressible information on StTW is available at $Re_\tau=200$, in our study the target value is set at the higher $Re_\tau = 400$. This choice brings in extra computational costs, but avoids issues with relaminarization, that is expected to become significant at lower $Re_\tau$ in view of the increased effectiveness of StTW in the compressible regime.

For each case (defined by a pair of values for $M_w^b$ and $Re_\tau$), two distinct simulations are carried out, which differ in the way the system is thermally managed. 
In one, dubbed Zero Bulk Cooling (ZBC), the bulk heating term $\Phi$ in Eq.\eqref{eq:energy} is set to zero, and the bulk temperature $T_b$ is left free to evolve until the aerodynamic heating rate and the heat flux at the wall are in balance. In the other, named Constrained Bulk Cooling (CBC), the heat produced within the flow is balanced not only by the wall heat flux, but also by a cooling source term $\Phi$ \citep{yu-xu-pirozzoli-2019}, which evolves to keep a constant $T_b$. 
A detailed description of the two strategies is provided later in \S\ref{subsec:comparison}, where the different implications of comparing at ZBC or CBC are discussed.


For each of the three values of $M_w^b$, a single uncontrolled and 42 cases with spanwise forcing are considered; each case is carried out twice, with ZBC and CBC.
Hence, the computational study consists of 258 simulations.
Table \ref{tab:parameters} summarizes the parameters for the 6 uncontrolled simulations.

\begin{table}
\centering
\begin{tabular}{ c c  c  c  c  c  c  c  c }
    & $M_w^b$ & $Re_\tau$ & $Re_b$ & $\Delta t^+$ & $N_x \times N_y \times N_z$ & $\Delta x^+$ & $\Delta y^+$ & $\Delta z^+$ \\ 
ZBC & 0.3 & 404 & 7115 & 0.007 & $768 \times 258 \times 528$ & 9.8 & 0.51--6.35 & 4.8	\\
ZBC & 0.8 & 400 & 6691 & 0.017 & $768 \times 258 \times 528$ & 9.8 & 0.51--6.28 & 4.8	\\
ZBC & 1.5 & 394 & 5751 & 0.025 & $1024 \times 258 \times 512$ & 7.4 & 0.50--6.19 & 4.9 \\
CBC & 0.3 & 403 & 7250 & 0.007 & $768 \times 258 \times 528$ & 9.8 & 0.51--6.35 & 4.8	\\
CBC & 0.8 & 399 & 7602 & 0.017 & $768 \times 258 \times 528$ & 9.8 & 0.51--6.28 & 4.8	\\
CBC & 1.5 & 387 & 8597 & 0.025 & $1024 \times 258 \times 512$ & 7.4 & 0.50--6.19 & 4.9 	\\
\end{tabular}
\caption{Parameters of the six uncontrolled simulations: Mach number $M_w^b$, friction Reynolds number $Re_\tau$, bulk Reynolds number $Re_b$, time step, mesh size and spatial resolution in each direction.}
\label{tab:parameters}
\end{table}

Periodic boundary conditions in the wall-parallel directions and no-slip and no-penetration conditions at the solid walls are applied for the velocity vector, and isothermal boundary conditions are used for the temperature.
In the cases with control, the no-slip condition for the spanwise velocity component is modified to apply the travelling wave \eqref{eq:sttw}.
The wave amplitude is fixed at $A^+ = 12$, and 42 different combinations of wavelength $\kappa_x^+$ and frequency $\omega^+$ are considered. 
Here and throughout the paper, the + superscript denotes quantities expressed in wall units of the uncontrolled case.

\begin{figure}
\centering
\includegraphics[width=0.8\textwidth]{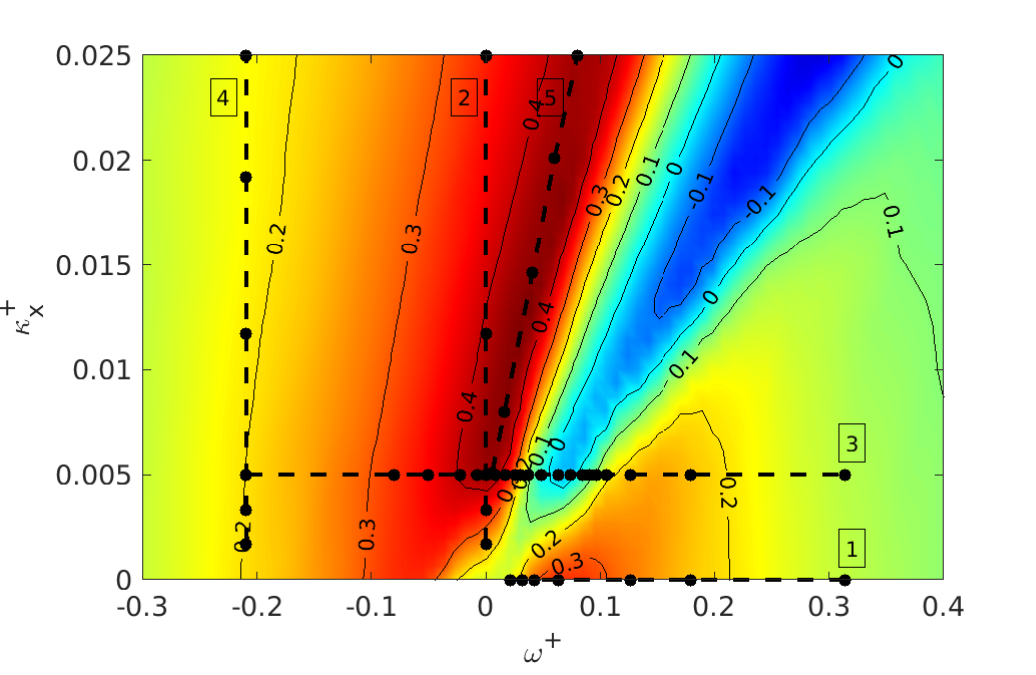}
\caption{Incompressible drag reduction versus $\kappa_x^+$ and $\omega^+$, at $A^+ = 12$ and $Re_\tau = 400$. The map is obtained from \cite{gatti-quadrio-2016} via interpolation of their datasets at $Re_\tau=200$ and $Re_\tau=1000$. The dots on the dashed lines correspond to the present compressible simulations.}
\label{fig:map}
\end{figure}

Figure \ref{fig:map} plots the incompressible drag reduction map, and dots identify the control parameters of the present simulations. 
The incompressible drag reduction map resembles the original one computed by \cite{quadrio-ricco-viotti-2009} at $Re_\tau=200$. Since the present value of $Re_\tau=400$ differs, this map is obtained via interpolation from the two datasets at $Re_\tau=200$ and $Re_\tau=1000$ produced by \cite{gatti-quadrio-2016} (see \S\ref{sec:results} for details). 
The simulations sample the parameter space along five lines, all visible in figure \ref{fig:map}.
In particular, the oscillating-wall case (dashed line 1 in figure \ref{fig:map}) at $\kappa_x^+=0$ is chosen to replicate data by \cite{yao-hussain-2019}, and sampled with 7 simulations (all with positive frequency, since negative frequencies at $\kappa_x=0$ can be obtained by symmetry). 
The steady wave at $\omega^+ = 0$ is scanned by 5 simulations along line 2; line 3 at constant $\kappa_x^+ = 0.005$ contains 20 points, crosses the low-$Re$ incompressible maximum drag reduction, and also cuts through the region of drag increase. Five simulations along line 4 explore the area of low drag reduction at large negative frequencies. Lastly, line 5 with 5 points analyses the ridge of maximum drag reduction. 

The size of the computational domain is ($L_x , L_y , L_z$) = ($6 \pi h, 2 h, 2\pi h$) in the streamwise, wall-normal and spanwise direction for the uncontrolled cases. For the controlled cases with $\kappa_x \ne 0$, $L_x$ is slightly adjusted on a case-by-case basis to fit the nearest integer multiple of the streamwise wavelength $\lambda_x$. In the case of longest forcing wavelength, two waves are contained by the computational domains.  

Although the discretization parameters have been chosen to replicate or improve those used in related studies, we have explicitly checked for the effect of wall-normal discretization and spanwise size of the computational domain. One specific case which yielded one of the largest drag reductions (namely the CBC case at $\kappa_x^+=0.005$ and $\omega^+=0.0251$) has been repeated by independently doubling $N_y$ and $L_z$. Starting from a baseline value for the friction coefficient of $C_f=3.41402 \times 10^{-3}$, we have measured $C_f=3.41347 \times 10^{-3}$ with doubled $N_y$ and $C_f=3.41733 \times 10^{-3}$ with doubled $L_z$. In both cases, the difference is below 0.1\%.

Statistics are computed with a temporal average of no less than $T_{ave} = 700 h/U_b$, after discarding the initial transient. The statistical time averaging error on the skin friction coefficient is estimated via the procedure introduced by \cite{russo-luchini-2017}. After propagating the error on the drag reduction, the corresponding uncertainties are found to be so small that the error bars are smaller than the symbols used in the figures in \S\ref{sec:results}.

\subsection{Performance indicators}
\label{subsec:performance}
The control performance is evaluated in terms of the dimensionless indicators drag reduction rate $\DR \%$, input power $P_{in} \%$ and net power saving $P_{net} \%$. 
These definitions, introduced by \cite{kasagi-hasegawa-fukagata-2009}, are suitable for CFR studies. 
The drag reduction rate describes the relative reduction of (dimensional) pumping power $P^*$ per unit channel area:
\begin{equation}
\DR \% = 100 \frac{P^*_0 - P^*}{P^*_0}
\label{eq:DR}
\end{equation}
where the subscript $0$ refers to the uncontrolled flow. Since all the simulations run at CFR, $\DR$ is equivalent to the reduction of the skin-friction coefficient $C_f = 2 \tau_w / (\rho_b U_b^2)$, and \eqref{eq:DR} can be expressed in terms of $C_f$ as:
\begin{equation}
\DR \%= 100 \left( 1 - \frac{C_f}{C_{f,0}} \right).
\label{eq:DRcf}
\end{equation}
The time-averaged pumping power per unit channel area is computed as: 
\begin{equation}
P^* = \frac{U_b}{T_{ave} L_x L_z} \int_{t_i}^{t_f} \int_{0}^{L_x} \int_{0}^{L_z}{\tau_x\,dx\,dz\,dt}
\label{eq:power}
\end{equation}
where $\tau_x$ is the streamwise component of the instantaneous wall-shear stress, and $T_{ave} = t_f-t_i$ is the interval for time averaging, defined by the final time $t_f$ and the time $t_i$ at which the initial transient is elapsed and a meaningful average can be taken. 
The control power $P_{c} \%$ is the power required to create the wall forcing while neglecting the losses of the actuation device, and is expressed as a fraction of the pumping power $P_0^*$. When the CBC strategy is employed, the power $P_\Phi$ required to cool the bulk flow should also be accounted for. Hence, the complete expression for the input power $P_{in}$ is:

\begin{align}
P_{in}\% = & P_{c}\% + P_\Phi\% = \notag \\
= & \frac{1}{P_0^*} \frac{100}{T_{ave} L_x L_z} \int_{t_i}^{t_f} \int_{0}^{L_x} \int_{0}^{L_z}{W\,\tau_z\,dx\,dz\,dt} +  \frac{100}{T_{ave}}
 \int_{t_i}^{t_f} \frac{\Phi}{\Phi_0^*} \,dt
\label{eq:Pin}
\end{align}
where $\tau_z$ is the spanwise component of the instantaneous wall-shear stress, $W$ the enforced spanwise wall velocity, and $\Phi_0^*$ the cooling power of the reference case.  
Finally, to compare benefits and costs of the control, the net energy saving rate $P_{net}$ is defined as:
\begin{equation}
P_{net} \%= \DR \%- P_{in} \%.
\label{eq:Pnet}
\end{equation}

\subsection{On the comparison strategy}
\label{subsec:comparison}
As mentioned above in \S\ref{subsec:parameters}, we consider two strategies to run the compressible channel flow, once $M_w^b$ and $Re_\tau$ are fixed.

The first one, denominated Zero Bulk Cooling (ZBC), sets to zero the bulk heating/cooling term $\Phi$ in Eq.\eqref{eq:energy}: the bulk temperature is thus free to increase until, at equilibrium, the heat produced within the flow is balanced by the heat flux at the walls. This setup corresponds to the one originally adopted by \cite{coleman-kim-moser-1995} for the plane channel, and employed in all previous compressible studies of drag reduction by spanwise wall motion \citep{fang-lu-shao-2009,yao-hussain-2019,ruby-foysi-2022}. 
ZBC simulations indicate that compressibility leads to larger drag reduction achieved by spanwise forcing. However, with ZBC the spanwise forcing causes $T_b$ to increase above the value of the uncontrolled flow, in a way that depends on the control parameters; the different heat transfer rates make it difficult to discern the specific effects of compressibility and wall cooling.
Furthermore, the equilibrium thermal condition achieved when the bulk temperature is free to evolve corresponds to extremely cold walls; the consequent large heat transfer rates are not representative of typical external flows, for which active techniques like spanwise forcing are primarily attractive.

To overcome these issues, a second strategy is considered, that is expected to provide more insight on the performance of flow control.
With this strategy, named Constrained Bulk Cooling (CBC), the heat produced within the flow is balanced not only by the heat flux through the walls, but also by a cooling source term $\Phi$, that is computed at each time step to keep the bulk temperature constant.

Following \cite{zhang-etal-2014}, we specify the thermal condition of the system by using the diabatic parameter $\Theta$, also named dimensionless temperature:
\begin{equation}
\Theta = \frac{T_w - T_b}{T_r - T_b},
\label{eq:Theta}
\end{equation}
where $T_r$ is the recovery temperature:
\begin{equation}
T_r = \left(1 + \frac{\gamma-1}{2} r \left( M^b_w \right)^2 \right) T_b ,
\label{eq:Tr}
\end{equation}
with $\gamma = c_p/c_v$ the heat capacity ratio, and $r$ the recovery factor, a coefficient that, according to \cite{shapiro-1953}, for a turbulent flow over a flat surface is $r= Pr^{1/3}$.

Recent studies \citep{cogo-etal-2023} have shown that a constant diabatic parameter, or equivalently a constant Eckert number \citep{wenzel-gibis-kloker-2022}, is the proper condition under which compressible flows at different Mach numbers should be compared.
The parameter $\Theta$ represents the fraction of the available kinetic energy transformed into thermal energy at the wall \citep{modesti-etal-2022}, and the importance of wall cooling increases when $\Theta$ decreases.
In this study we set $\Theta = 0.75$, which corresponds to a moderately cold wall.

\begin{figure}
\centering
\includegraphics[width=\textwidth]{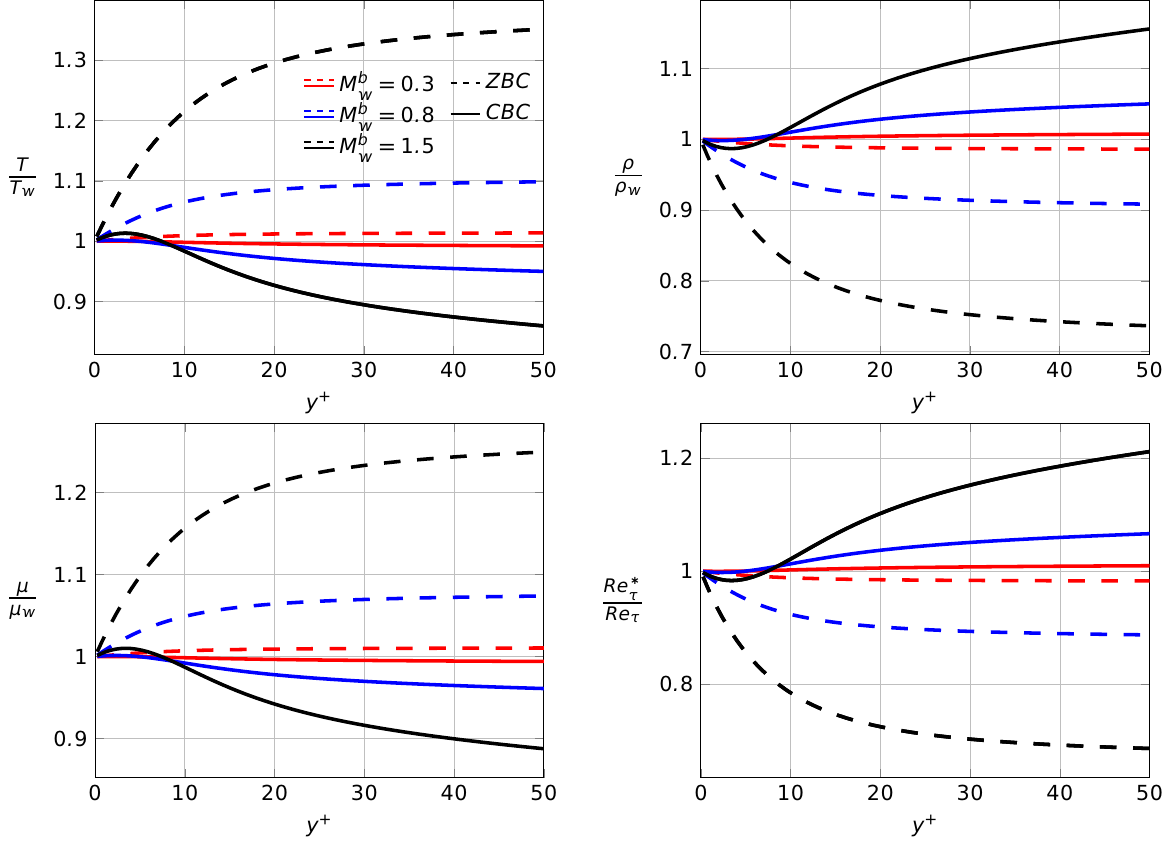}
\caption{Temperature (top left), density (top right), dynamic viscosity (bottom left) and semi-local Reynolds number (bottom right) profiles in the wall region of a canonical compressible channel flow at $M_w^b=0.3, 0.8$ and $1.5$, with ZBC (dashed lines) and CBC (continuous lines).}
\label{fig:re}
\end{figure}

The main differences arising from the two channel configurations, ZBC and CBC, can be appreciated in figure \ref{fig:re}, where temperature, density and dynamic viscosity profiles across the channel are shown for the uncontrolled flow cases.
In ZBC, at equilibrium the mean temperature profile monotonically increases from its minimum at the wall to its maximum at the channel centreline; the same trend is shared by the viscosity, whereas the opposite trend is observed for the density.
Since $T_b$ grows with $M_w^b$, the profile of $T/T_w$ across the channel, shown in the top left panel of figure \ref{fig:re}, gets progressively steeper at the wall with increasing $M_w^b$.
While $T/T_w \approx 1$ for the subsonic $M$, at the channel centre for $M_w^b=1.5$ (not shown) the mean temperature is about 39\% higher than at the wall.
The significant changes (especially for $M_w^b = 1.5$) of thermodynamic properties across the buffer layer imply that the local properties are quite different from the wall properties. 
In particular, the friction-velocity based Reynolds number $Re_\tau$ is intended to be constant across the comparison while $M_w^b$ varies. However, in the buffer layer the semi-local Reynolds number $Re^*_\tau=Re_\tau \sqrt{(\rho \mu_w)/(\rho_w \mu)}$ \citep{huang-coleman-bradshaw-1995} is far from constant (see bottom right panel of \ref{fig:re}), and varies significantly as a function of $M_w^b$. 

With CBC, instead, $Re_\tau^*$ across the channel is such that its value in the buffer layer is still similar to the one at the wall (with a maximum observed increase of 2\% for $M_w^b=1.5$ at $y^+=10$) with a variation of less than 1.5\% around the mean value of $Re_\tau^*$ at $y^+=10$, for the three values of $M_w^b$. 
Moreover, the profile of $T/T_w$ across the channel qualitatively resembles the temperature distribution of a typical compressible boundary layer.
In fact, at supersonic speeds the wall temperature can be considered for practical purposes to be very close to the recovery temperature of the flow, implying a very low heat exchange at the wall. Smaller values of $\Theta$ imply a cooler wall, and a local maximum of $T/T_w$ further from the wall.
For $\Theta=0.75$, the local peak is minor and located right within the buffer layer, as shown in the top left panel of figure \ref{fig:re}.

\begin{figure}
\centering
\includegraphics[trim={200 0 200 0},width=0.49\textwidth]{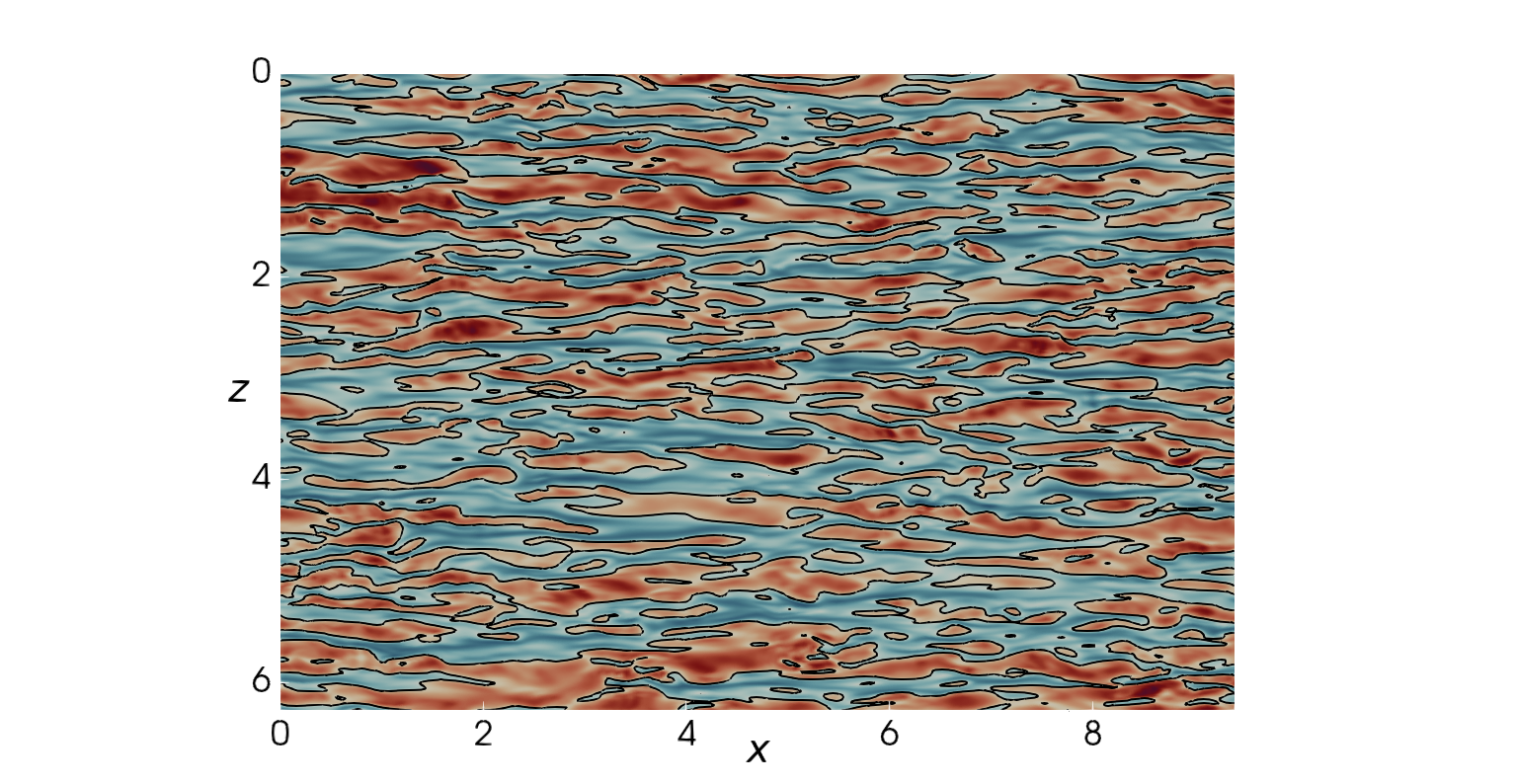} 
\includegraphics[trim={100 0 100 0},width=0.498\textwidth]{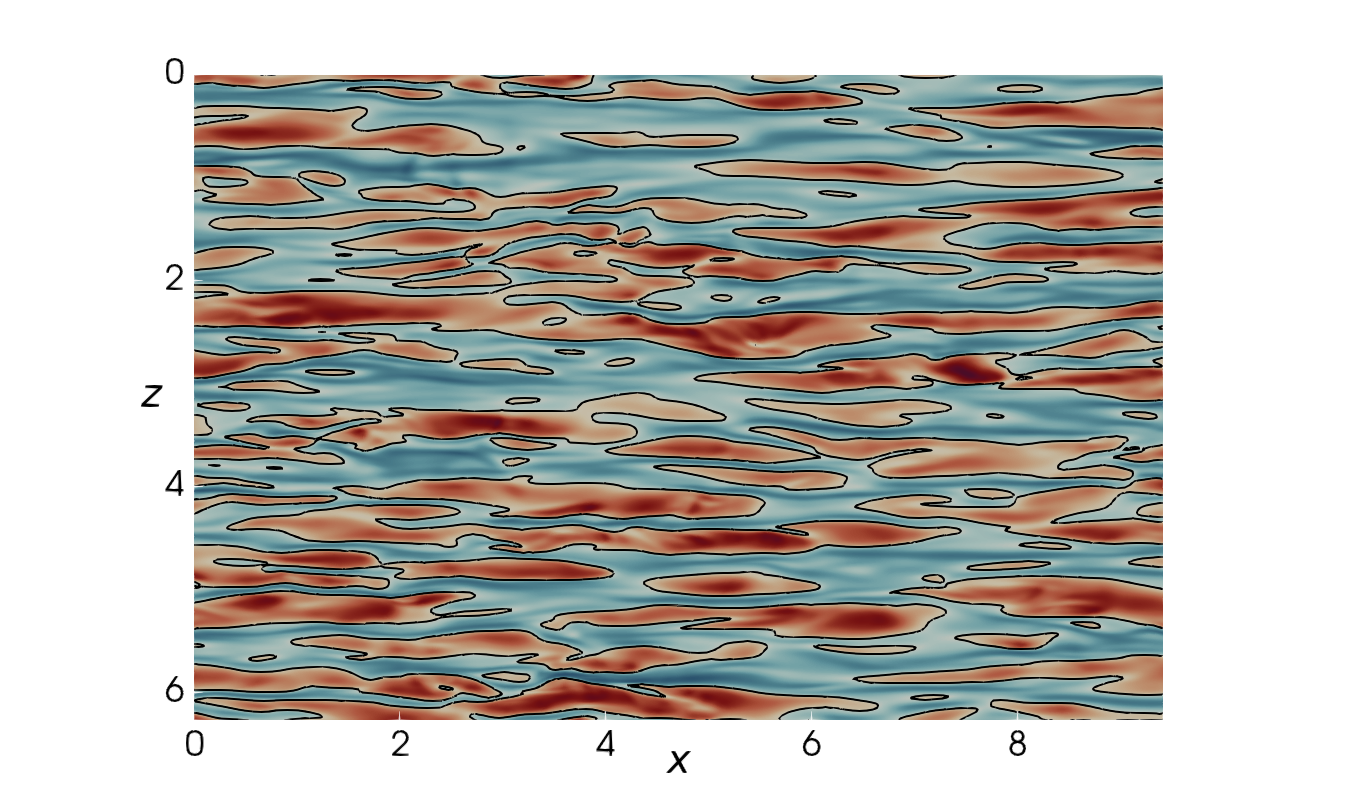}
\includegraphics[trim={200 0 200 0},width=0.49\textwidth]{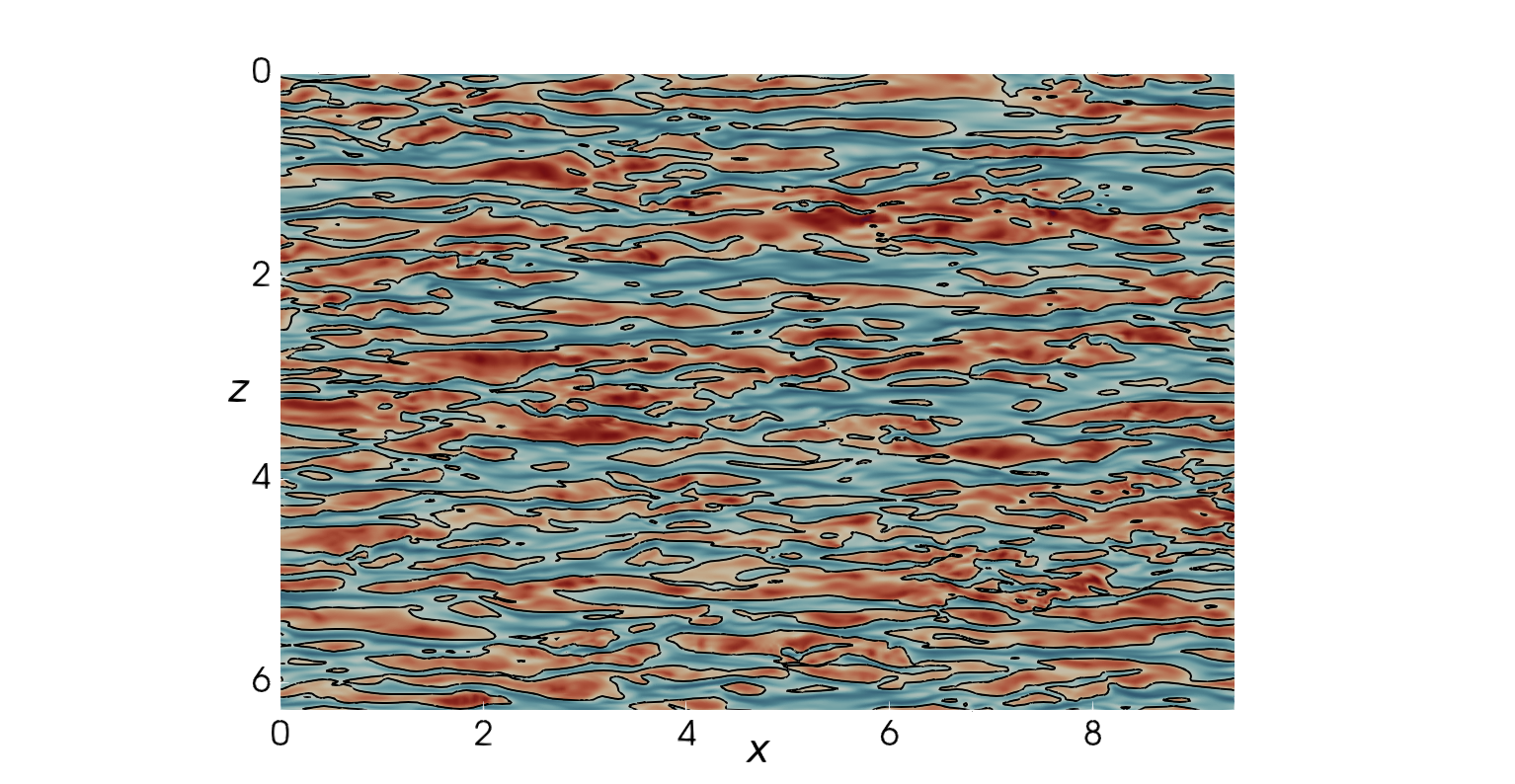}
\includegraphics[trim={100 0 100 0},width=0.498\textwidth]{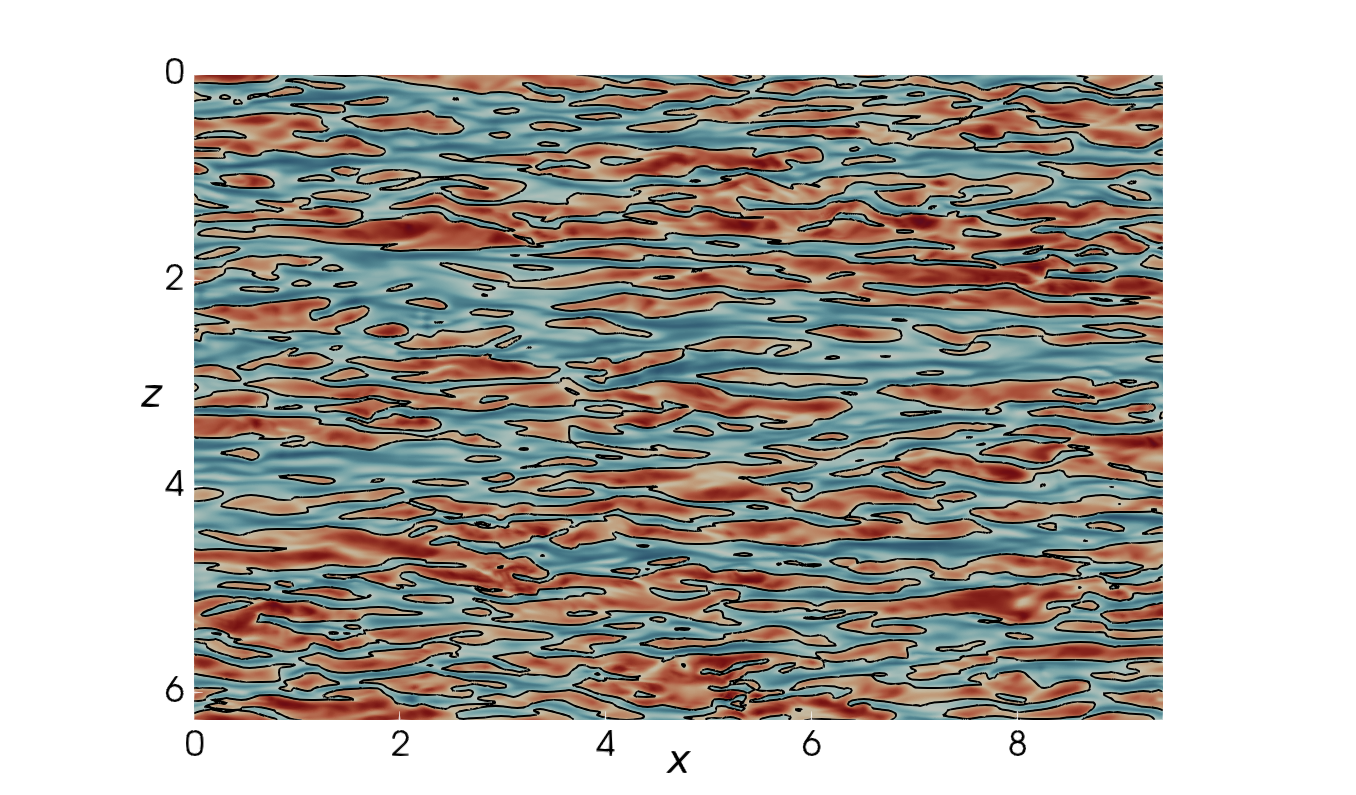}
\caption{Streamwise velocity fluctuations $u^+$ in a wall-parallel portion of the $x-z$ plane at $y^+=10$ for ZBC (top) and CBC (bottom) at $M_w^b=0.3$ (left) and $M_w^b=1.5$ (right) for the uncontrolled case. The blue-to-red colorscale ranges from $-10$ to $+10$; the black line is for the zero contour level.}
\label{fig:streaks}
\end{figure}

The difference between ZBC and CBC can be visually appreciated by looking at the near-wall turbulent structures in the uncontrolled flow, shown in figure \ref{fig:streaks}.
It is known \citep{coleman-kim-moser-1995} that by increasing $M_w^b$ the low-velocity streaks become longer, less wavy and more widely spaced. This is indeed confirmed in the top row of figure \ref{fig:streaks}, where color contours of an instantaneous field of streamwise velocity fluctuations computed with ZBC at $y^+=10$ is plotted for $M_w^b=0.3$ (left) and $M_w^b=1.5$ (right). 
However, when switching to CBC (bottom row), the streaks appear not to differ significantly between the subsonic and the supersonic cases. This suggests that a matching diabatic parameter allows to discriminate those changes of the near-wall structures that directly derive from compressibility effects from those linked to a change in the wall-normal temperature profile.  
In fact, a non-uniform temperature across the channel implies changes to other thermodynamic properties (i.e. density and viscosity), and their wall values are no more fully representative of the physics in the buffer layer. 
This observation is essential when the purpose of the study is to assess skin-friction drag reduction by spanwise forcing, whose physical mechanism is not fully uncovered yet, but certainly resides within the thin transversal Stokes layer which interacts with the near-wall cycle occurring in the buffer layer. 
When the actuation parameters scale in viscous wall units, their effects in the buffer layer are not easily comparable in the ZBC case.

\begin{figure}
\centering
\includegraphics[width=1\textwidth]{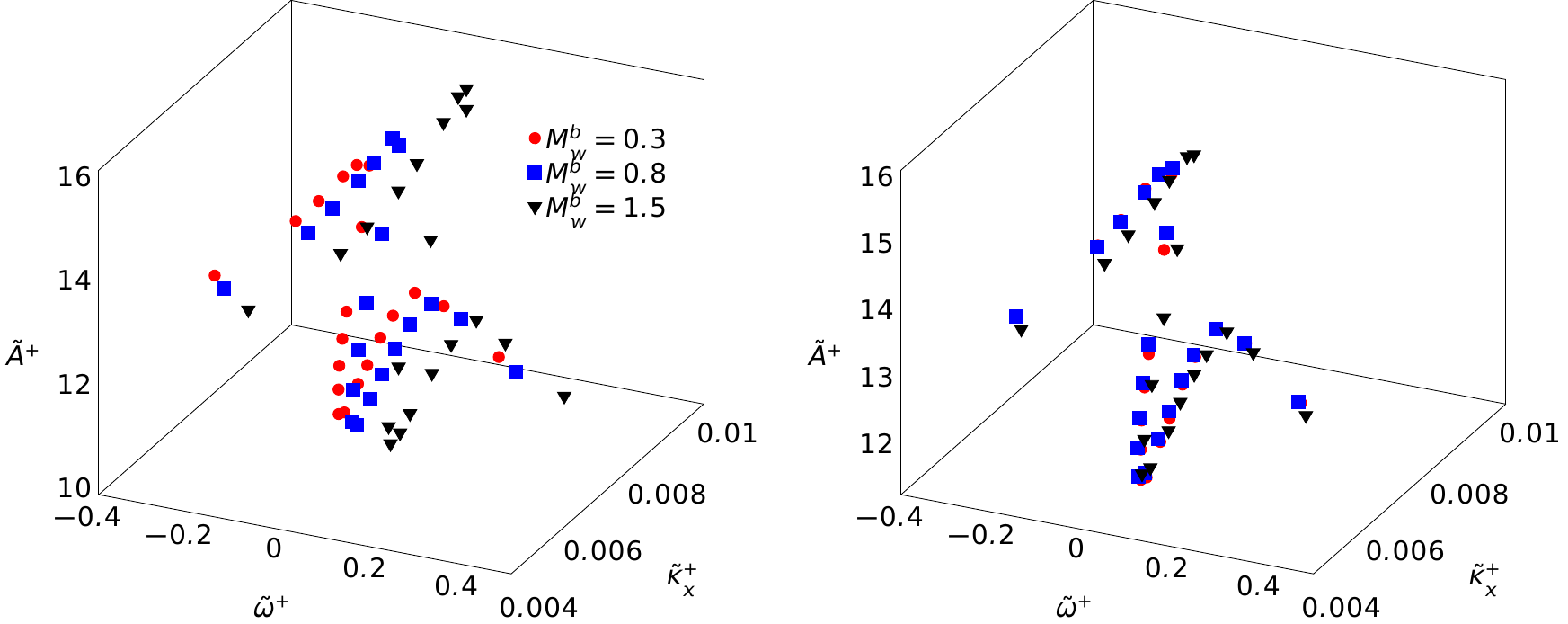} 
\caption{Frequency $\tilde{\omega}^{+}$, wavenumber $\tilde{\kappa}_x^+$ and amplitude $\tilde{A}^+$ of the control forcing for the travelling waves at $\kappa_x^+=0.005$ (line 3 of figure \ref{fig:map}) made dimensionless with the thermodynamic properties of the actuated flow at $y^+=10$.}
\label{fig:parameters}
\end{figure}

As an example, figure \ref{fig:parameters} plots the control parameters $\tilde{\omega}^+$, $\tilde{\kappa}^+_x$ and $\tilde{A}^+$ of the simulations taken along line 3 of figure \ref{fig:map}. The parameters are still scaled in wall units, but the reference quantities are taken at $y^+=10$, i.e. made dimensionless with viscous units built with density and viscosity measured for the actuated flow at $y^+=10$, for the ZCB (left) and CBC (right) comparison strategy.
Figure \ref{fig:parameters} is effective at showing that with ZBC the buffer layer experiences a forcing whose set of parameters changes with the Mach number, whereas with CBC the simulation parameters match at the various $M_w^b$, and enable the comparison of compressibility effects for a given control.

\section{Drag reduction and power savings}
\label{sec:results}

The database produced in the present work is used for a comprehensive analysis of the effect of compressibility on the drag reduction and power budget performance of StTW.
The reference Reynolds number of choice is $Re_\tau=400$, i.e. higher than $Re_\tau=200$, where most of the incompressible information is available, to avoid full or partial relaminarization.
Data at $Re_\tau=400$ are also relatively free from the low-$Re$ effects that plague results obtained at $Re_\tau=200$. 
Obviously, the downsides are a larger computational cost, and a limited number of incompressible data to directly compare with. Results at $M_w^b=0.3$ are compared to those of \cite{hurst-yang-chung-2014} for the oscillating wall, stationary waves and the travelling waves at fixed wavenumber. For the oscillating wall, a few data points from \cite{ricco-quadrio-2008} are also available. 
For the other control cases, the main incompressible comparison data are the StTW results of \cite{gatti-quadrio-2016}.
Their comprehensive datasets at $Re_\tau=200$ and $Re_\tau=1000$, available as Supplementary Material to their paper, are interpolated to obtain drag reduction for arbitrary combinations of the control parameters. As suggested in that paper, drag reduction data is expressed in terms of the vertical shift $\Delta B^+$ of the streamwise mean velocity profile in its logarithmic region, which minimizes the effect of the small computational domain and reduces the $Re$ effect on $\DR$. 
In fact, $\Delta B^+$ becomes a $Re$-independent measure of drag reduction, once $Re$ is sufficiently large (they tentatively suggested $Re_\tau > 2000$) for the mean profile to feature a well-defined logarithmic layer. 
Since $\Delta B^+$ is still $Re$-dependent at the present values of $Re$, we interpolate linearly the $\Delta B^+$ data by \cite{gatti-quadrio-2016} between $Re_\tau=200$ and $Re_\tau=1000$ to retrieve $\Delta B^+$ at $Re_\tau = 400$. 
Note that, owing to the small computational domain, the $Re_\tau=200$ data by \cite{gatti-quadrio-2016} slightly overestimate drag reduction, particularly at small frequencies and wavelengths. 
The incompressible control power is interpolated at $Re_\tau=400$ from data of \cite{gatti-quadrio-2016}, by assuming a power law dependence with $Re_\tau$, as stated by \cite{ricco-quadrio-2008} and \cite{gatti-quadrio-2013}.

The few available compressible data are from \cite{yao-hussain-2019}, who considered the oscillating wall only, at the slightly higher $Re_\tau=466$ for $M_w^b=0.8$ and $Re_\tau=506$ for $M_w^b=1.5$. Moreover, the datapoints computed by \cite{ruby-foysi-2022} for a stationary wave are at $M_w^b=0.3, Re_\tau=396$ and $M_w^b=1.5, Re_\tau=604$.

\subsection{Drag reduction}
\label{sec:DR}


\begin{figure}
\centering
\includegraphics[width=\textwidth]{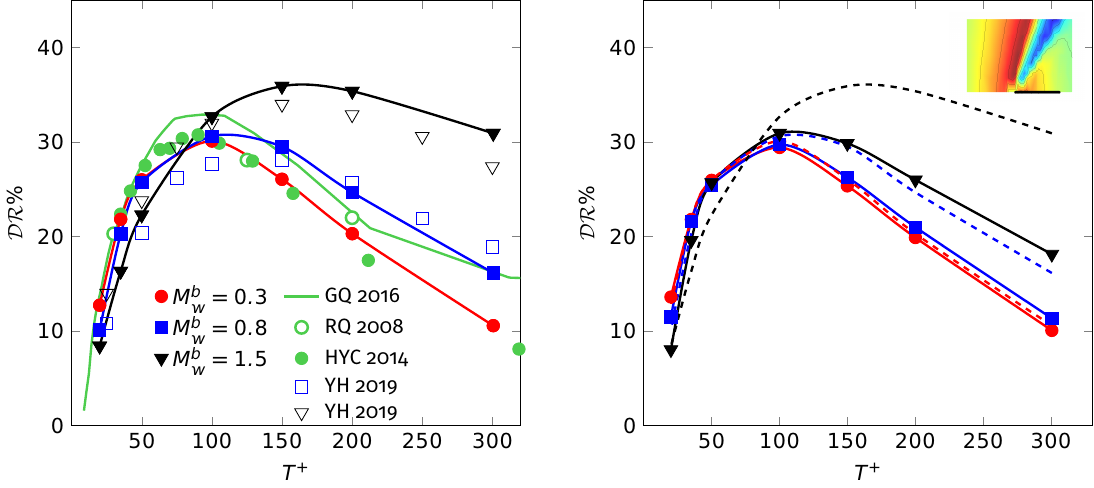}
\caption{Drag reduction rate versus oscillation period $T^+$ for the oscillating wall (line 1 of figure \ref{fig:map}, see inset), for ZBC (left) and CBC (right). Incompressible data are in green: solid line without symbols from \cite{gatti-quadrio-2016}, solid symbols from \cite{hurst-yang-chung-2014}, and open symbols from \cite{ricco-quadrio-2008}. The blue and black open symbols are from \cite{yao-hussain-2019} at $M_w^b=0.8, Re_\tau=466$ and $M_w^b=1.5, Re_\tau=506$. Solid lines indicate interpolation. Dashed lines on the right panel are results for ZBC. 
}
\label{fig:DR-line1}
\end{figure}

Figure \ref{fig:DR-line1} shows the drag reduction rate obtained for the temporally oscillating wall, i.e. along line 1 of figure \ref{fig:map}, as a function of the oscillating period $T^+$. 

We first consider the ZBC case on the left. For $M_w^b=0.3$, $\DR$ grows with $T^+$ up to a maximum at about $T^+=100$, and then monotonically shrinks. This is in agreement with the incompressible results of \cite{hurst-yang-chung-2014}, \cite{ricco-quadrio-2008} and \cite{gatti-quadrio-2016}, whose interpolated data, as expected, slightly overpredict $\DR$, especially at large periods. 
This is due to the combined effect of low $Re$ and small computational domain employed in that study, which -- particularly for the oscillating wall -- leads to partial relaminarization during the cycle.
The curves at higher $M_w^b$ are qualitatively similar, but tend to remain below the incompressible data at small periods, and to go above them at large ones. 
Near the optimal period, compressibility makes the maximum $\DR \%$ grow, and shift towards larger periods: for $M_w^b=0.3$ the maximum drag reduction is $\DR_{0.3}^m= 30.3 \%$ at $T^+=100$, whereas $\DR_{0.8}^m = 30.6 \%$ at $T^+=100$, and for $M_w^b=1.5$ it becomes $\DR_{1.5}^m= 35.9 \%$ at $T^+=150$. 
This picture confirms the compressible results at $Re_\tau=200$ discussed by \cite{yao-hussain-2019}, except for the supersonic case, where they reported a monotonic increase of $\DR \%$ with $T^+$. 
This is ascribed to the partial relaminarization occurring at $Re_\tau=200$ when drag reduction is large; the present study, owing to its higher $Re_\tau=400$, is able to identify a well defined $\DR \%$ peak even in the supersonic regime. 
Figure \ref{fig:DR-line1} also includes results at higher $Re_\tau$ from \cite{yao-hussain-2019} for the transonic and supersonic cases. Again, qualitative agreement is observed; quantitative differences are due to their slightly different Reynolds number, which is $Re_\tau=466$ for $M_w^b=0.8$ and $Re_\tau=506$ for $M_w^b=1.5$.

The right panel of figure \ref{fig:DR-line1} plots the results computed under CBC, and compares them with those under ZBC. 
The $M_w^b=0.3$ cases are almost identical; at this low $M_w^b$ compressibility effects are minor, and the difference between ZBC and CBC negligible. At larger $M_w^b$, however, with CBC the results show a much better collapse over the three values of $M_w^b$. The maximum drag reduction consistently occurs at $T^+=100$, and is nearly unchanged across the three cases.

Overall, the favorable effect of compressibility in terms of maximum drag reduction of the oscillating wall is confirmed. However, the significant increase of the maximum drag reduction reported by \cite{yao-hussain-2019} is only confirmed when the comparison is carried out with ZBC, whereas for CBC this increment is very limited.

\begin{figure}
\centering
\includegraphics[width=\textwidth]{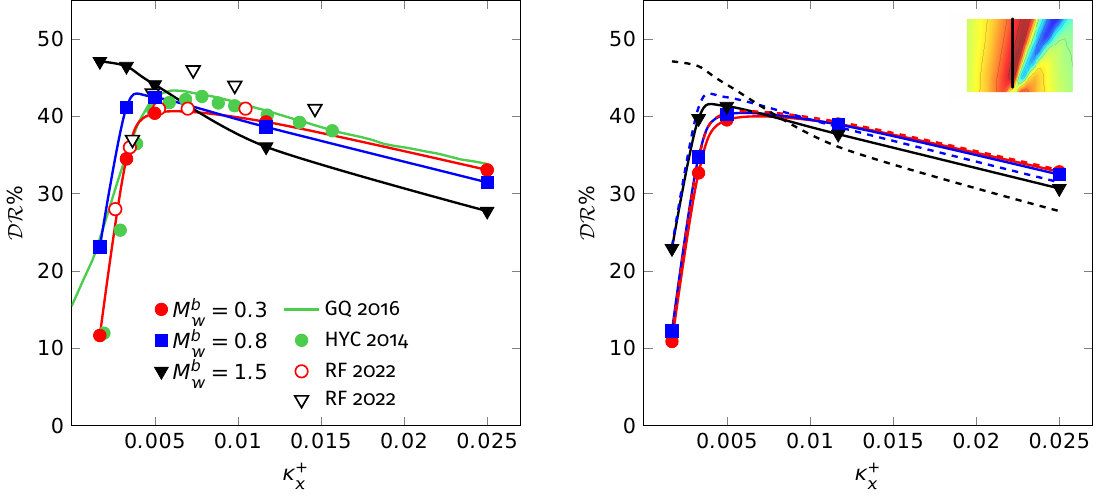}
\caption{Drag reduction rate versus wavenumber $\kappa_x^+$ for the steady waves (line 2 of figure \ref{fig:map}, see inset), for ZBC (left) and CBC (right). Incompressible data are in green and dashed lines are for ZBC, as in figure \ref{fig:DR-line1}. Red and black open symbols are from \cite{ruby-foysi-2022} at $M_w^b=0.3, Re_\tau=396$ and $M_w^b=1.5, Re_\tau=604$.} 
\label{fig:DR-line2}
\end{figure}

Figure \ref{fig:DR-line2} shows results for the stationary waves, i.e. along line 2 of figure \ref{fig:map}, plotted as a function of the streamwise wavenumber $\kappa_x$. 
The trend resembles that of the temporal oscillation. Again, at $M_w^b=0.3$ differences from the incompressible limit are minor. Once $M_w^b$ grows, a significant dependency on the wavenumber is observed: at large $\kappa_x$ $\DR \%$ slightly decreases, but at small  $\kappa_x$ it increases significantly.

For the ZBC dataset (left), a significant shift of the $\DR \%$ peak towards smaller wavenumbers is observed, with a peak value of $\DR_{0.3}^m = 40.4 \%$ for $\kappa_x^+=0.005$,  $\DR_{0.8}^m = 42.5 \%$ for $\kappa^+_x=0.005$, and $\DR_{1.5}^m = 47.1 \%$ for $\kappa_x^+=0.0017$. 
However, once the CBC comparison is considered (right), the overshoot at small $\kappa_x^+$ disappears; data at $M_w^b=0.3$ and $M_w^b=0.8$ collapse, and the supersonic case still presents its maximum at $\kappa^+_x=0.005$. 

Open symbols in the left panel of figure \ref{fig:DR-line2} are the results of \cite{ruby-foysi-2022}, computed with ZBC. One immediately notices their different trend compared to the present data.  
In fact, in their numerical experiments the value of the semi-local Reynolds number evaluated at the centreline was kept fixed at $Re^*_{\tau,c} = 400$: this implies a variation of $Re_\tau$ between $396$ and $604$ while moving from the subsonic to the supersonic case. In the present simulations, instead, $Re_\tau \approx 400$ at all $M$.
Additionally, in their study the forcing wavelength was scaled with semi-local quantities, so that a direct comparison is problematic. 
Red and black open symbols represent their results at $M_w^b=0.3$ and $M_w^b=1.5$, rescaled in viscous units: these rescaled data present the same trend observed here with CBC, with the supersonic case lacking the $\DR \%$ peak at the smallest $\kappa_x^+$, and suggest a qualitative similarity between a comparison based on a semi-local scaling and the present CBC strategy. 

\begin{figure}
\centering
\includegraphics[width=\textwidth]{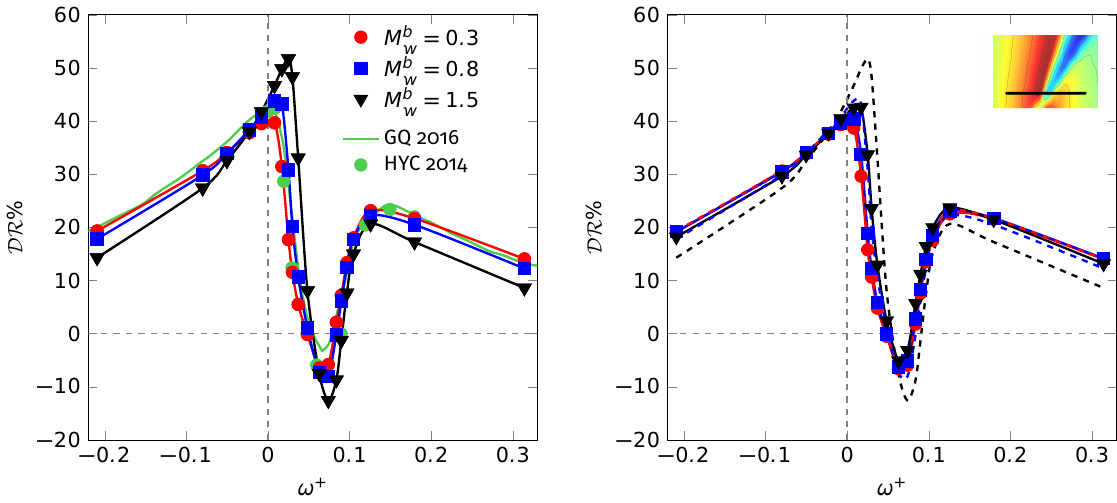}
\caption{Drag reduction rate versus frequency $\omega^+$ for the streamwise-travelling waves at $\kappa_x^+ = 0.005$ (line 3 of figure \ref{fig:map}, see inset), for ZBC (left) and CBC (right). Incompressible data are in green and dashed lines data are for ZBC, as in figure \ref{fig:DR-line1}.}
\label{fig:DR-line3}
\end{figure}

We now move on to consider a travelling wave, and plot in figure \ref{fig:DR-line3} how $\DR \%$ varies as a function of the frequency $\omega^+$ for a travelling wave at fixed $\kappa_x^+ = 0.005$, i.e. along line 3 of figure \ref{fig:map}.
Once again, data for $M_w^b=0.3$ do not differ from the incompressible ones.
At higher $M_w^b$, with ZBC the maximum drag reduction increases above the incompressible value, but, far from the peak, drag reduction levels are generally lower. 
The boost in maximum drag reduction grows with $M_w^b$, and is accompanied by a slight shift towards higher frequencies. At $M_w^b=1.5$, the peak is at $\omega^+=0.025$, and reaches the outstanding value of $\DR_{1.5}^m = 51.6 \%$. 
Increasing $M_w^b$ also intensifies the drag increase in the range $0.05 \lesssim \kappa_x^+ \lesssim 0.1$, with a maximum of 12.2\% for $M_w^b=1.5$.

Once again, if the comparison is carried out with the CBC criterion, the compressibility effects remain generally favourable, but become much smaller. The extra gain is extremely small, and the curves at varying $M_w^b$ nearly collapse. 

\begin{figure}
\centering
\includegraphics[width=\textwidth]{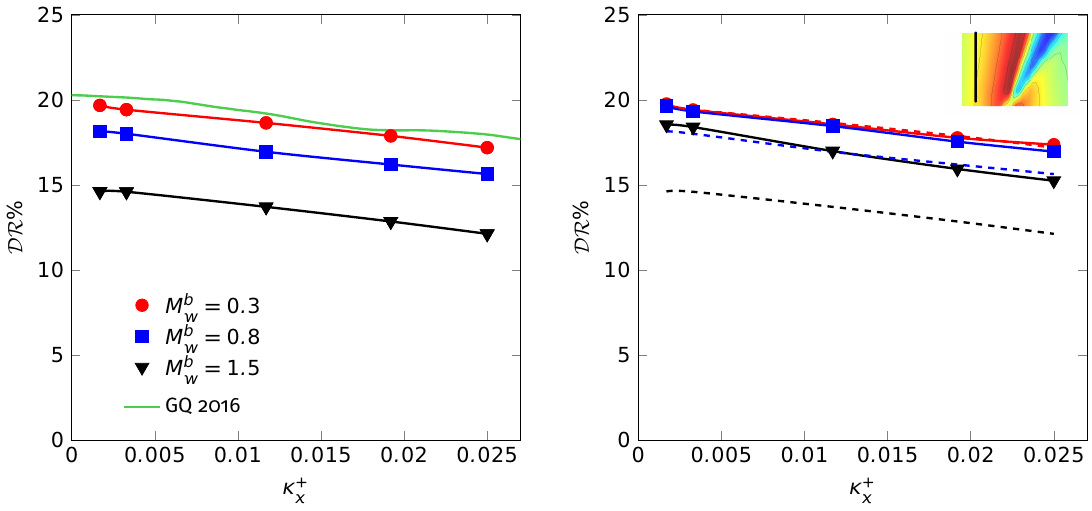}
\caption{Drag reduction rate versus wavenumber $\kappa_x^+$ for the travelling waves at $\omega^+ = -0.21$ (line 4 of figure \ref{fig:map}, see inset), for ZBC (left) and CBC (right). Incompressible data are in green, and dashed lines are for ZBC, as in figure \ref{fig:DR-line1}.}
\label{fig:DR-line4}
\end{figure}

Figure \ref{fig:DR-line4} reports the results computed for the points on the vertical line 4 of figure \ref{fig:map} at fixed $\omega^+ = -0.21$, where the incompressible $\DR \%$ is nearly constant with $\kappa_x^+$.
As for lines 1 and 3, compressibility is found to deteriorate the control performances at large (positive and negative) frequencies. 
However, this is emphasized by the ZBC comparisons, whereas CBC results show a much better collapse.

\begin{figure}
\centering
\includegraphics[width=\textwidth]{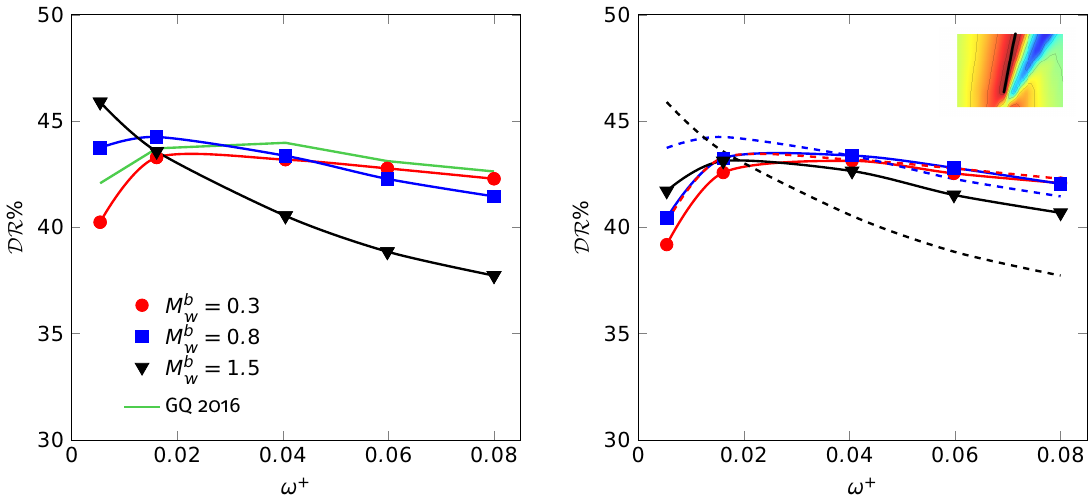}
\caption{Drag reduction rate versus frequency $\omega^+$ for the travelling waves for the optimal ridge (line 5 of figure \ref{fig:map}, see inset), for ZBC (left) and CBC (right). Incompressible data are in green, and dashed lines are for ZBC, as in figure \ref{fig:DR-line1}.}
\label{fig:DR-line5}
\end{figure}

Finally, results from simulations on line 5 in figure \ref{fig:map}, drawn along the ridge of optimal $\DR \%$ in the ($\omega-\kappa_x$) plane of parameters, are depicted in figure \ref{fig:DR-line5}.
It is worth recalling that, according to \cite{gatti-quadrio-2016}, this ridge and in particular its portion near the origin of the plane is where the largest changes with $Re$ are expected. 
Indeed, the subsonic points do not fully overlap with incompressible data, which inherit the low-$Re$ nature of the reference through the interpolation, and show a rather uniform value of $\DR \%$. The supersonic data lie below the subsonic ones at large frequencies, but outperform them at small frequencies. 
Once CBC is used, the collapse of the curves at different $M_w^b$ improves significantly, while the general changes remain qualitatively the same.

\subsection{Power budgets}
\label{sec:power}
Since StTW is an active form of flow control, quantifying the energy consumption of the control system is key to assess the overall efficiency: one needs to compare costs, i.e. the control energy, and benefits, i.e. the energy savings made possible by a reduction of the skin-friction drag. 

\begin{figure}
\centering
\includegraphics[width=\textwidth]{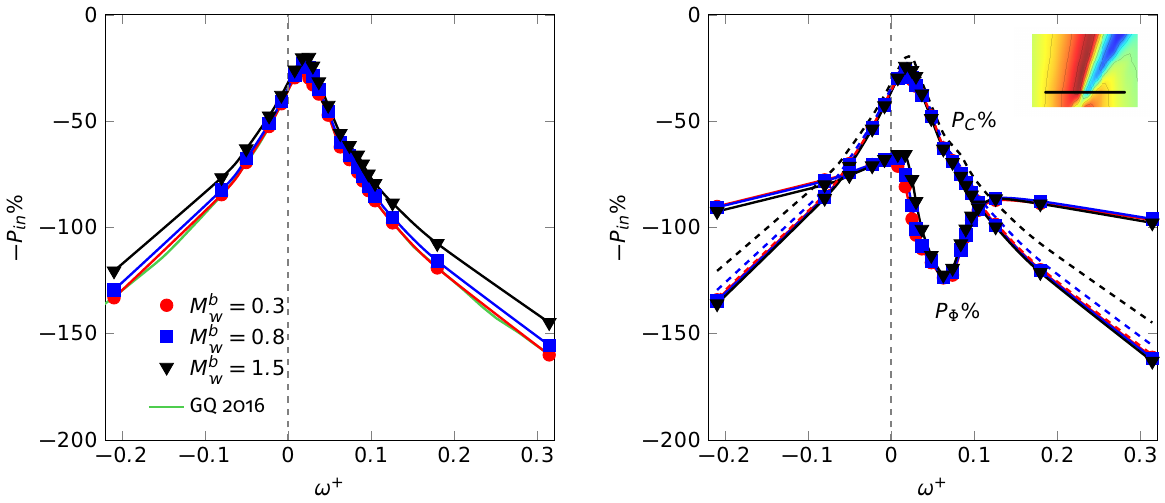}
\caption{Input power for the travelling waves with $\kappa_x^+ = 0.005$ (line 3 of figure \ref{fig:map}, see inset) for ZBC (left) and CBC (right). For CBC the two contributions to $P_{in} \%$, i.e. the control power $P_c \%$ and the cooling power $P_\Phi \%$ are plotted separately. Incompressible data are in green, and dashed lines are for ZBC.}
\label{fig:Pin}
\end{figure}

Figure \ref{fig:Pin} plots, as one example, the input power $P_{in} \%$ on line 3 of figure \ref{fig:map}. A similar scenario holds in the entire plane. For the ZBC comparison (left panel), the input power, which depends significantly on the control parameters, shows a decrease (in absolute value) with $M_w^b$, especially at large frequencies. 
With CBC, $P_{in} \%$ features two contributions: the control power and the cooling power. They turn out to be roughly of the same order of magnitude, and both have a minor dependence on $M_w^b$, yet the dependence of the latter on control parameters resembles the one of $\DR \%$. 
The extra cost to cool the flow is an effect of the additional term in the energy equation, which serves the purpose of yielding an internal flow with a temperature profile that resembles an external flow. In a true external flow, however, cooling would occur naturally: $P_{in} \%$ would reduce to the control power $P_c \%$. Since the control contribution to $P_{in} \%$ in StTW is a rather simple quantity that can be analytically predicted under the hypothesis of a laminar generalized Stokes layer \citep{quadrio-ricco-2011}, the perfect collapse of $P_c \%$ under CBC witnesses how the controlled cases are being properly compared.

\begin{figure}
\centering
\includegraphics[width=\textwidth]{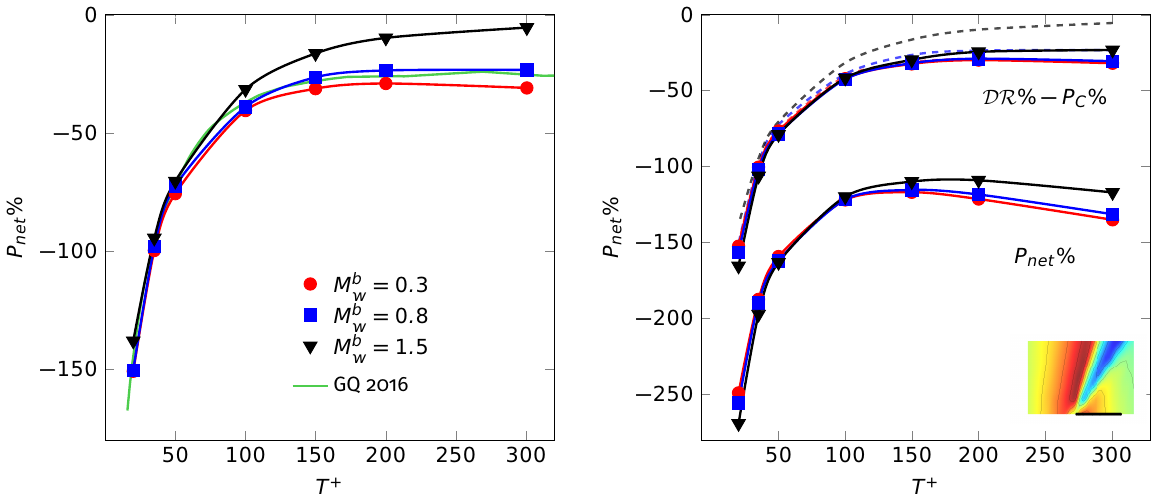}
\caption{Net power saving for the oscillating wall (line 1 of figure \ref{fig:map}, see inset), for ZBC (left) and CBC (right). Incompressible data are in green. The right panel also plots $\DR \% - P_c \%$ (top set of curves), where dashed lines are for ZBC.}
\label{fig:P-line1}
\end{figure}

Figure \ref{fig:P-line1} plots the net power saving $P_{net} \%$ for the temporal wall oscillations, i.e. along line 1 of figure \ref{fig:map}. 
The left panel is computed with ZBC; in agreement with the incompressible case, for $A^+=12$ no net saving is obtained. However, the power budget improves with the Mach number, and at $M_w^b=1.5$ it approaches zero. This is due to the combined effect of increasing $\DR \%$ (for $T^+ \gtrsim 100$, see figure \ref{fig:DR-line1}), and decreasing $P_{in} \%$ (especially for small $T$). 
The right panel of figure \ref{fig:P-line1} plots $P_{net} \%$ under CBC (lower set of curves), and the net power saving without accounting for the cooling power, namely $\DR \% - P_c \%$. 
Since $P_c \%$ and $P_\Phi \%$ are of the same order of magnitude, $P_{net} \%$ becomes largely negative: the interesting outcome of the ZBC case vanishes.  
However, when only $P_c \%$ is considered, $P_{net} \%$ becomes comparable with the ZBC case (upper set of curves), albeit the positive compressibility effect decreases substantially. 

\begin{figure}
\centering
\includegraphics[scale=0.7]{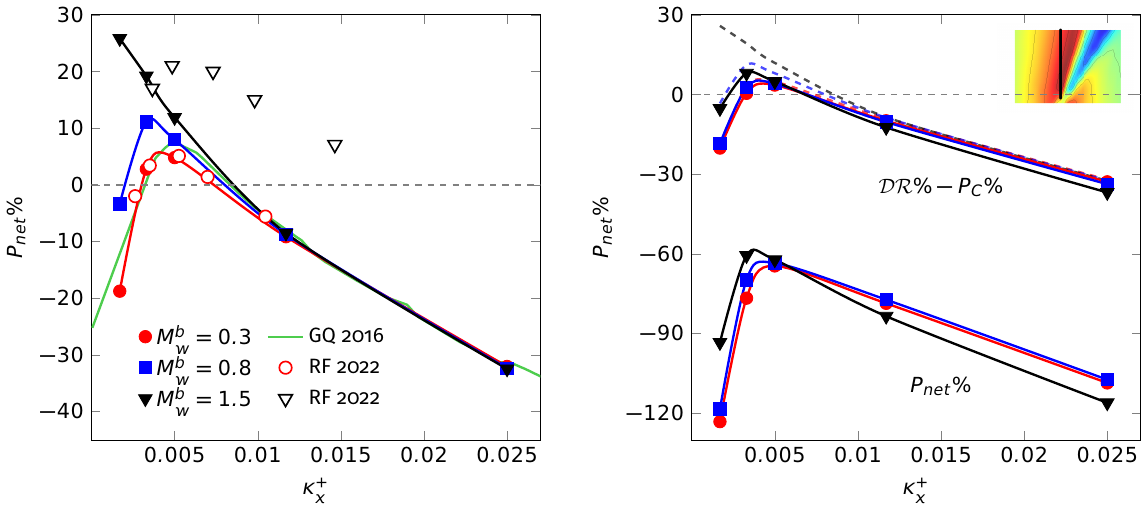}
\caption{Net power saving for the steady waves (line 2 of figure \ref{fig:map}, see inset), for ZBC (left) and CBC (right). Incompressible data are in green. The right panel also plots $DR \% - P_c \%$ (top set of curves) where dashed lines are for ZBC. Red and black open symbols are from \cite{ruby-foysi-2022} at $M_w^b=0.3, Re_\tau=396$ and $M_w^b=1.5, Re_\tau=604$.}
\label{fig:P-line2}
\end{figure}

Examining data along line 2 of figure \ref{fig:map} (stationary waves), which passes near the absolute maximum of drag reduction, is instructive. The plot is shown in figure \ref{fig:P-line2}. For a ZBC comparison (left), the net saving increases substantially with $M_w^b$ for $\kappa_x^+ < 0.012$, such that the maximum shows a 5-fold increase, from 5\% in the incompressible case to 25.8\% for $M_w^b=1.5$ 
The peak is also observed to shift towards smaller $\kappa_x^+$. 
Under CBC, however, much of the improvement disappears, and the curves almost collapse, with only a small residual effect for the supersonic curve. When $P_{net} \%$ takes into account the cooling power, the outcome is negative regardless of the control parameters.

Results from \cite{ruby-foysi-2022} at ZBC and at fixed $Re^*_{\tau,c}$ are also plotted in the left panel of figure \ref{fig:P-line2}. They are computed at rather small wavenumbers, and overlap to the present data for $M_w^b=0.3$, but indicate much larger savings at $M_w^b=1.5$. 
Nevertheless, their trend resembles the one obtained here at CBC, and indicate the presence of a local maximum, and the lack of explosive savings at vanishing wavenumbers. 

\begin{figure}
\centering
\includegraphics[width=\textwidth]{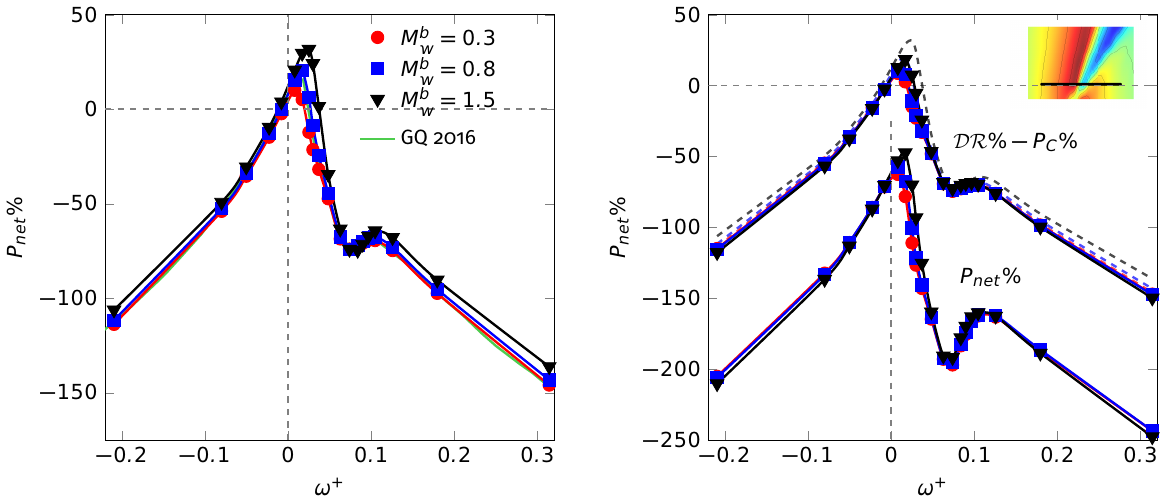}
\caption{Net power saving for the travelling waves with $\kappa_x^+ = 0.005$ (line 3 of figure \ref{fig:map}, see inset) for ZBC (left) and CBC (right). Incompressible data are in green. The right panel also plots $DR \% - P_c \%$ (top set of curves) where dashed lines are for ZBC.}
\label{fig:P-line3}
\end{figure}

Figure \ref{fig:P-line3} plots the net power saving for travelling waves at fixed $\kappa_x^+=0.005$ (line 3 of figure \ref{fig:map}).
The ZBC comparison shows a large increase of $P_{net} \%$, up to 31.4 \% for the largest $M$; the peaks shift towards larger positive $\omega$. 
Interestingly, the peaks of $\DR \%$ and $P_{in} \%$ occur around the same frequency, and they are both enhanced by compressibility.
When the comparison is carried out at CBC, however, once again the curves show a tendency to overlap, and the maximum saving shrinks to 17.8\% for $\DR \% - P_{in} \%$, which remains an interesting figure, but in line with the incompressible case. If both contributions to $P_{in} \%$ are included, $P_{net} \%$ is largely negative at every $\omega$.

\section{Concluding discussion}
\label{sec:conclusions}

We have studied how spanwise forcing implemented via streamwise-travelling waves of spanwise velocity at the wall alters the skin-friction drag in compressible flows.
A set of 258 direct numerical simulations for a turbulent plane channel flow are carried out, for subsonic ($M_w^b=0.3$), transonic ($M_w^b=0.8$) and supersonic ($M_w^b=1.5$) speeds, at the baseline friction Reynolds number of $Re_\tau=400$. 
The available literature information, which includes only few such studies for compressible flows, is significantly extended; in particular, travelling waves are considered here for the first time. 
The study considers the control performance for the temporally oscillating wall ($\kappa_x=0$), the steady wave ($\omega=0$), travelling waves at fixed wavenumber $\kappa^+_x=0.005$ and at fixed frequency $\omega^+ = -0.21$, and the ridge of maximum drag reduction corresponding to waves travelling with a slow forward speed. 
All the simulations are run by keeping the bulk velocity constant in time as well as between unforced and forced cases. 

Besides the bulk velocity, in the compressible setting a further quantity related to the energy equation must be kept constant to enable a proper comparison.
Since its choice impacts the qualitative outcome of the study, we employ and compare two different strategies.
The first, that we indicate with Zero Bulk Cooling or ZBC, is commonly used for duct flows, and lets the bulk temperature evolve freely until an asymptotic value is reached at which the heat produced within the flow is balanced by the heat flux through the isothermal walls. 
Unfortunately, ZBC leads to different bulk temperatures for each simulation, and in the present context it hinders the physical interpretation of results. 

In a second approach, named Constrained Bulk Cooling or CBC, the value of the bulk temperature is kept constant during the simulations, by means of a bulk cooling term in the energy equation. To do so, the value of the diabatic parameter $\Theta$ is fixed across both the values of the Mach number and the control parameters of the StTW, implying that a fixed portion of bulk flow kinetic energy is converted into thermal energy, and that extra energy is spent for the cooling process.
Using the diabatic parameter (or, equivalently, the Eckert number) has been recently considered by \cite{cogo-etal-2023} as a means to achieve a similar wall cooling across different values of the Mach number. Extending a $\Theta$-based comparison to account for different values of $\Theta$ with flow control and drag reduction is an interesting future development of the present study.

\begin{figure}
\centering
\includegraphics[width=0.5\textwidth]{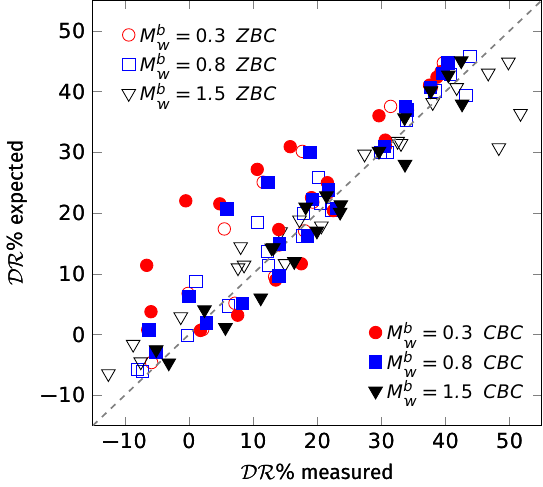} 
\caption{Drag reduction for the streamwise-travelling waves at $\kappa_x^+ = 0.005$ measured in the compressible regime versus drag reduction of the incompressible regime when the control parameters are scaled with the thermodynamic properties of each different case at $y^+=10$.}
\label{fig:diag}
\end{figure}

Results of the simulations show that StTW remain fully effective in transonic and supersonic flows, thus extending available results for the oscillating wall and the steady waves. 
In fact, drag reduction can be higher in compressible flows than in incompressible ones, when frequency and wavenumber of the forcing are small. 
However, the improvement appears to be substantial only when the comparison is carried out at ZBC. When CBC is used, only marginal improvements are detected; curves at various $M_w^b$ tend to collapse and to replicate the incompressible behaviour.
Figure \ref{fig:diag} shows for the controlled flow at $\kappa_x^+=0.005$ (line 3 of the map of figure \ref{fig:map}) the drag reduction measured by the simulations of the present work plotted against the drag reduction of the incompressible case. The control parameters are made dimensionless with the thermodynamic properties of each case at $y^+=10$ (see \S\ref{subsec:comparison}). 
Most points lie on the diagonal line: drag reduction becomes constant with the Mach number, once the effect of the changed thermodynamics is removed. The few outliers are points of the map where drag reduction gradients are extremely large, and the limited number of available incompressible data leads to a poor interpolation, as already pointed out in \S\ref{sec:results}.
This picture demonstrates that, once spurious thermodynamic changes are factored out, compressibility has little to no effect on the drag reduction performance of the travelling waves.

Similar results hold for the power budget: StTW yield large net energy savings, even in the compressible regime, but the impressive improvements observed with ZBC against the incompressible reference do not carry over to the CBC comparison, which broadly replicates the incompressible results. The last statement is only valid as long as the extra cooling power implied by CBC is neglected, on the basis that it represents an artefact to obtain an internal flow with a temperature profile that resembles that of an external flow.

Hence, choosing the comparison strategy is key to properly describe how drag reduction and power savings of an active drag reduction technique change in the compressible regime.
In a way, this reminds of the incompressible case, where early studies for the oscillating wall claimed ``disruption of turbulence" only because comparing at the same bulk velocity implies an important reduction of $Re_\tau$ when drag reduction is achieved. 
While ZBC is certainly apt to describe internal flows, the observed drag reduction figures are significantly larger than their incompressible counterpart primarily because the control parameters affect the terms of the comparison. 
A CBC comparison, in which the dimensionless temperature remains constant with $M$ and across the controlled cases, seems more appropriate, and in fact yields data that overlap well when the Mach number is varied. With CBC, only a small, albeit non negligible, extra drag reduction and net power saving are found in comparison to the incompressible case. 


\section*{Acknowledgments}
Computing resources have been provided by CINECA on the computer Marconi100, under the ISCRA B projects DWSWBLI and SPADRACO. 

\section*{Funding} 
This research received no specific grant from any funding agency, commercial or not-for-profit sectors.

\section*{Declaration of Interests} 
The authors report no conflict of interest.

\bibliographystyle{jfm}

\end{document}